\documentclass[twoside,twocolumn,9pt]{article}
\usepackage{extsizes}
\usepackage[super,sort&compress,comma]{natbib} 
\usepackage[version=3]{mhchem}
\usepackage[left=1.5cm, right=1.5cm, top=1.785cm, bottom=2.0cm]{geometry}
\usepackage{balance}
\usepackage{mathptmx}
\usepackage{sectsty}
\usepackage{graphicx} 
\usepackage{lastpage}
\usepackage[format=plain,justification=justified,singlelinecheck=false,font={stretch=1.125,small,sf},labelfont=bf,labelsep=space]{caption}
\usepackage{float}
\usepackage{fancyhdr}
\usepackage{fnpos}
\usepackage[english]{babel}
\addto{\captionsenglish}{%
  
}
\usepackage{array}
\usepackage{droidsans}
\usepackage{charter}
\usepackage[T1]{fontenc}
\usepackage[usenames,dvipsnames]{xcolor}
\usepackage{setspace}
\usepackage[compact]{titlesec}
\usepackage{hyperref}
\usepackage{bm}
\usepackage{xcolor}
\hypersetup{
    colorlinks=true,
    linkcolor=blue,
    citecolor=blue,
    urlcolor=cyan,
}

\usepackage{epstopdf}

\usepackage{soul} 

\usepackage{caption}

\DeclareCaptionLabelFormat{adja-page}{\hrulefill\\#1 #2 \emph{(previous page)}}

\definecolor{cream}{RGB}{222,217,201}

\begin{document}
\pagestyle{fancy}
\thispagestyle{plain}
\fancypagestyle{plain}{
\renewcommand{\headrulewidth}{0pt}
}

\makeFNbottom
\makeatletter
\renewcommand\LARGE{\@setfontsize\LARGE{15pt}{17}}
\renewcommand\Large{\@setfontsize\Large{12pt}{14}}
\renewcommand\large{\@setfontsize\large{10pt}{12}}
\renewcommand\footnotesize{\@setfontsize\footnotesize{7pt}{10}}
\makeatother

\renewcommand{\thefootnote}{\fnsymbol{footnote}}
\renewcommand\footnoterule{\vspace*{1pt}%
\color{cream}\hrule width 3.5in height 0.4pt \color{black}\vspace*{5pt}} 
\setcounter{secnumdepth}{5}

\makeatletter 
\renewcommand\@biblabel[1]{#1}            
\renewcommand\@makefntext[1]%
{\noindent\makebox[0pt][r]{\@thefnmark\,}#1}
\makeatother 
\renewcommand{\figurename}{\small{Fig.}~}
\sectionfont{\sffamily\Large}
\subsectionfont{\normalsize}
\subsubsectionfont{\bf}
\setstretch{1.125} 
\setlength{\skip\footins}{0.8cm}
\setlength{\footnotesep}{0.25cm}
\setlength{\jot}{10pt}
\titlespacing*{\section}{0pt}{4pt}{4pt}
\titlespacing*{\subsection}{0pt}{15pt}{1pt}

\fancyfoot{}
\fancyfoot[LO,RE]{\vspace{-7.1pt}\includegraphics[height=9pt]{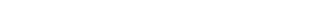}}
\fancyfoot[CO]
{\vspace{-7.1pt}\hspace{13.2cm}\includegraphics{head_foot/LF}}
\fancyfoot[CE]{\vspace{-7.2pt}\hspace{-14.2cm}\includegraphics{head_foot/LF}}
\fancyfoot[RO]{\footnotesize{\sffamily{1--\pageref{LastPage} ~\textbar  \hspace{2pt}\thepage}}}
\fancyfoot[LE]{\footnotesize{\sffamily{\thepage~\textbar\hspace{3.45cm} 1--\pageref{LastPage}}}}
\fancyhead{}
\renewcommand{\headrulewidth}{0pt} 
\renewcommand{\footrulewidth}{0pt}
\setlength{\arrayrulewidth}{1pt}
\setlength{\columnsep}{6.5mm}
\setlength\bibsep{1pt}

\makeatletter 
\newlength{\figrulesep} 
\setlength{\figrulesep}{0.5\textfloatsep} 

\newcommand{\topfigrule}{\vspace*{-1pt}%
\noindent{\color{cream}\rule[-\figrulesep]{\columnwidth}{1.5pt}} }

\newcommand{\botfigrule}{\vspace*{-2pt}%
\noindent{\color{cream}\rule[\figrulesep]{\columnwidth}{1.5pt}} }

\newcommand{\dblfigrule}{\vspace*{-1pt}%
\noindent{\color{cream}\rule[-\figrulesep]{\textwidth}{1.5pt}} }

\makeatother

\twocolumn[
  \begin{@twocolumnfalse}
{\includegraphics[height=30pt]{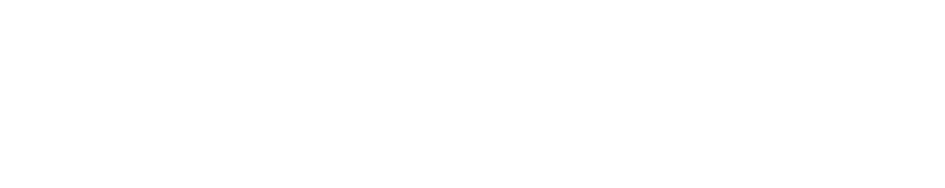}\hfill\raisebox{0pt}[0pt][0pt]{\includegraphics[height=55pt]
{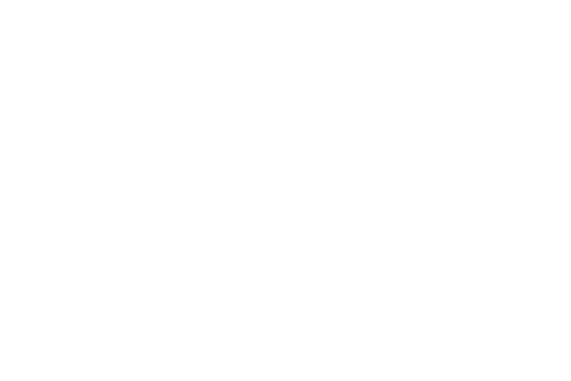}}\\[1ex]
\includegraphics[width=18.5cm]
{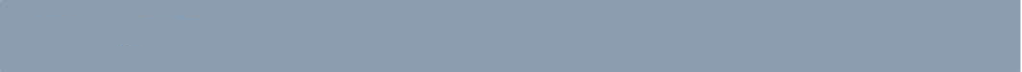}}\par
\vspace{1em}
\sffamily
\begin{tabular}{m{4.5cm} p{13.5cm} }

\includegraphics{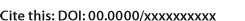} & \noindent\LARGE{\textbf{Optical heterodyne microscopy of operating spin Hall nano-oscillator arrays}} \\
\vspace{0.3cm} & \vspace{0.3cm} \\

 & \noindent\large{A. Alem\'an,\textit{$^{a}$} A. A. Awad,\textit{$^{abc\dag}$} S. Muralidhar\textit{$^{a}$}, 
 R. Khymyn\textit{$^{a}$},
 A. Kumar,\textit{$^{abc}$} A. Houshang,\textit{$^{a}$} D. Hanstorp,\textit{$^{d}$} and J. \AA kerman\textit{$^{abc}$}} \\

\includegraphics{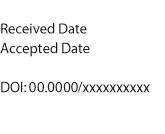} & \noindent\normalsize{Optical heterodyne detection is a powerful technique for characterizing a wide range of physical excitations. Here, we use two types of optical heterodyne detection techniques (fundamental and parametric pumping) to microscopically characterize the high-frequency auto-oscillations of single and multiple nano-constriction spin Hall nano-oscillators (SHNOs). To validate the technique and demonstrate its robustness, we study SHNOs made from two different material stacks, NiFe/Pt and W/CoFeB/MgO, and investigate the influence of both the RF injection power and the laser power on the measurements, comparing the optical results to conventional electrical measurements. To demonstrate the key features of direct, non-invasive, submicron, spatial, and phase-resolved characterization of the SHNO magnetodynamics, we map out the auto-oscillation magnitude and phase of two phase-binarized SHNOs used in Ising Machines. This proof-of-concept platform establishes a strong foundation for further extensions, contributing to the ongoing development of crucial characterization techniques for emerging computing technologies based on spintronics devices.} \\

\end{tabular}

 \end{@twocolumnfalse} \vspace{0.6cm}

  ]

\renewcommand*\rmdefault{bch}\normalfont\upshape
\rmfamily
\section*{}
\vspace{-1cm}


\footnotetext{\textit{$^{a}$~Applied Spintronics Group, Department of Physics, University of Gothenburg, Gothenburg 412 96, Sweden}}
\footnotetext{\textit{$^{b}$~Center for Science and Innovation in Spintronics, Tohoku University, 2-1-1 Katahira, Aoba-ku, Sendai 980-8577 Japan}}
\footnotetext{\textit{$^{c}$~Research Institute of Electrical Communication, Tohoku University, 2-1-1 Katahira, Aoba-ku, Sendai 980-8577 Japan}}
\footnotetext{\textit{$^{d}$~Department of Physics, University of Gothenburg, 412 96 Gothenburg, Sweden}}
\footnotetext{\textit{$^{\dag}$~ahmad.awad@gu.se}}




Spin torque and spin Hall nano-oscillators (STNOs and SHNOs) \cite{chen2016ieee} are promising candidates in the emerging era of neuromorphic computing \cite{grollier2020natelc, Torrejon2017Nature,Zahedinejad2020Two-dimensionalComputing,Zahedinejad2022natmat}. The latter show particular promise thanks to easy fabrication and miniaturization to nano-scopic dimensions \cite{durrenfeld2017nanoscale, haidar2019ntcom}, compatibility with CMOS technology \cite{Zahedinejad2018apl, ranjbar2014ieeeml}, room temperature operation \cite{Mazraati2016apl, mazraati2018apl}, scalability \cite{Awad2020apl}, and strong interactions and mutual synchronization in both long chains \cite{Awad2016natphys,Kumar2023RobustChains} and two-dimensional (2D) arrays \cite{Zahedinejad2020Two-dimensionalComputing}. Recently it was shown that parametrically pumped 2D SHNO networks can exhibit phase binarization, allowing them to operate as oscillator-based Ising machines solving combinatorial optimization problems \cite{Albertsson2021UltrafastNano-oscillators, Houshang2022Phase-BinarizedMachines}. 
This promising approach is an alternative to quantum computing, overcoming some of its challenges, such as cryogenic operating temperatures, large operating facilities, and kilowatt power consumption \cite{Boixo2014EvidenceQubits,Hamerly2019ExperimentalAnnealer}. 

In the rapidly evolving landscape of emerging spintronic computation technologies, optical tools offer a versatile and powerful alternative to traditional electrical characterization. Techniques such as Brillouin Light Scattering (BLS) \cite{Demokritov2008Micro-brillouinNanostructures,Hache2019CombinedNano-oscillators}, and Time-Resolved Magneto-Optical Kerr Effect (TR-MOKE) microscopy \cite{Ogasawara2023Time-ResolvedMicroscopy, Buess2005ExcitationsVortex, Spicer2018TimeNano-oscillator} have paved the way for noninvasive, high-resolution local imaging of magnetic dynamics down to sub-micrometer scales. Optothermal control of SHNOs \cite{Muralidhar2022OptothermalNano-oscillators} has also demonstrated the additional potential of direct optical manipulation, eliminating the complexities associated with physical connections.

However, the adoption of optical tools poses its own set of challenges. Although BLS microscopy has been the go-to choice in magnonics \cite{Sebastian2015Micro-focusedNanoscale}, it has several drawbacks, such as large size, high cost, and slow measurement speeds. Overcoming these limitations requires optical tools that are fast, compact, cost-effective, and easy to implement. Frequency-resolved MOKE (FR-MOKE) is a novel optical heterodyne approach that offers a route to fulfill the above requirements. This technique is an optical counterpart to conventional ferromagnetic resonance spectrometers, replacing one end of the physical circuit (sample to detector) with an optical link, and has previously demonstrated effectiveness in imaging passive spin dynamics in nanomagnet arrays \cite{Schneider2007SpinEffect,Shaw2009SpinDamping} and studying antenna-generated propagating spinwaves in thin films \cite{Shiota2020ImagingDetection,Liensberger2019Spin-waveEffect}. 

Expanding the applications of MOKE-based optical heterodyne detection, here we evaluate its value for the characterization of active SHNO devices and arrays. In contrast to conventional electrical power spectral density (PSD) measurements of SHNOs, where a DC stimulus drives the free-running magnetization precession and a spectrum analyzer captures the resulting microwave voltage, the optical heterodyne approach requires a vector network analyzer (VNA) with an additional RF stimulus to injection lock the SHNO to a reference signal. Similar to phase-resolved BLS microscopy, the optical heterodyne method, hence, does not capture the truly free-running properties of the SHNO. It is therefore important to investigate the impact of the injected signal. As we will see, by studying the behavior of the microwave signal \emph{vs.}~the power of the injected RF stimulus and extrapolating to zero injection, it is still possible to extract the intrinsic linewidth of the SHNO. The impact of the laser power on the measurement should also be considered. We, therefore, carry out a detailed examination of the influence of both the RF and the laser power, comparing our optical results with electrical measurements to establish the reliability of this method for characterizing single and multiple SHNO devices. In SHNO arrays, optical probing allows the investigation of individual oscillators, showcased here by spatially profiling a synchronized 2-SHNO system. Our approach features faster data acquisition and a streamlined setup, opening avenues for embedded applications and the potential for miniaturization. This positions the proposed method as a promising direction for advancing computing technologies based on SHNO networks ~\cite{Zahedinejad2020Two-dimensionalComputing,Houshang2022Phase-BinarizedMachines}.

\begin{figure}
    \centering
    \includegraphics[width=0.9\linewidth]{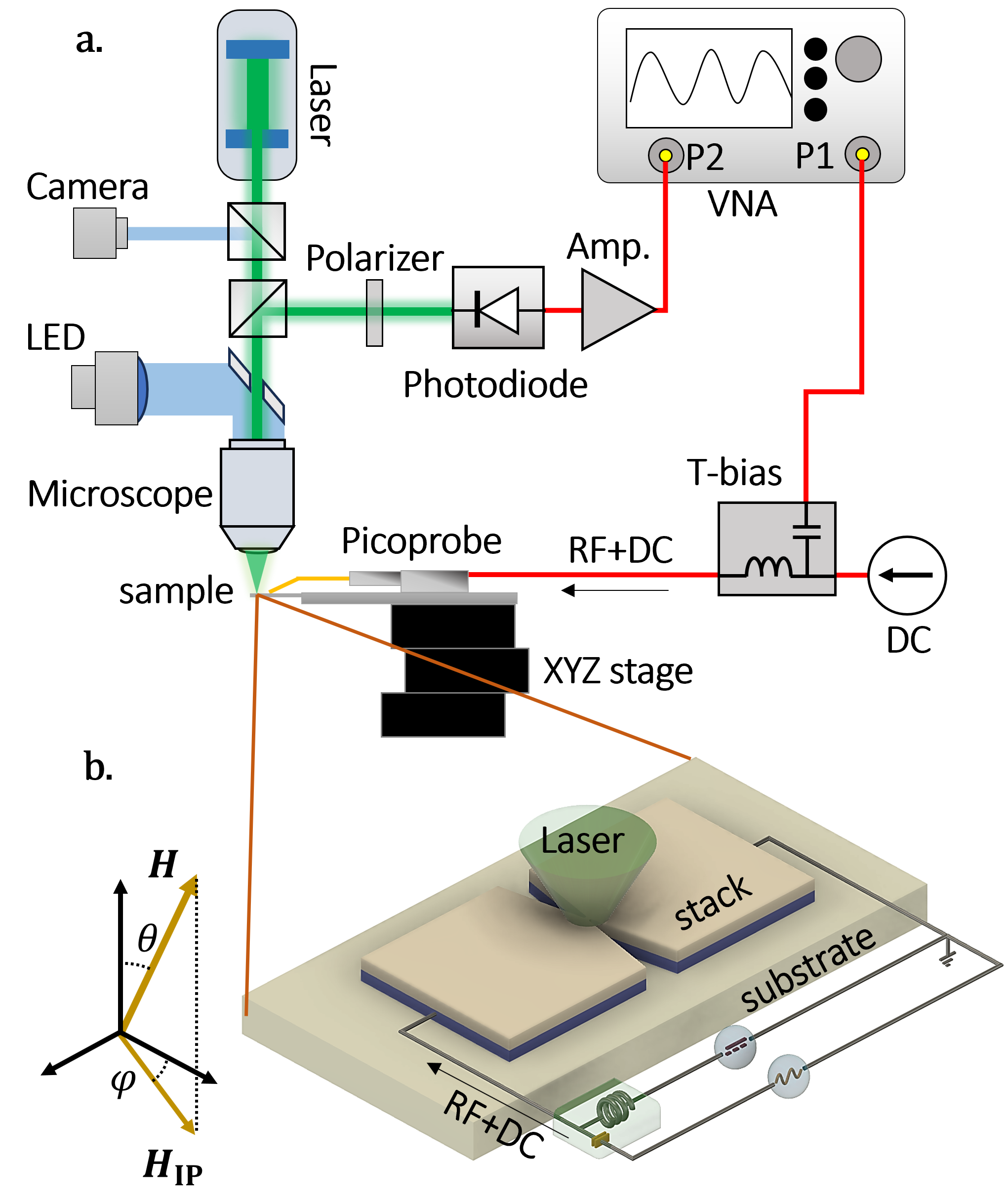}
    \caption{a. Schematic of the experimental setup:
    Red lines represent electrical connections. Green lines represent the laser path. In blue, the LED illumination light path.
    b. Zoom view of the sample (substrate and stack) and reference axes. The SHNO is subject to an external DC+RF stimulus. The external magnetic field $H$, its polar ($\theta$) and azimuthal ($\phi$) field angles, and its in-plane component $\mathit{H_{IP}}$ are shown. The incident laser probe is shown in green.}
    \label{fig:1 setup}
\end{figure}

To validate the generality of optical heterodyne detection of magnetic auto-oscillations, we first investigated two distinct SHNO material systems---W/CoFeB/MgO and NiFe/Pt---and then interacting SHNOs of W/CoFeB. The first material system consists of a W (5 nm)/CoFeB (1.4 nm)/MgO (2 nm) trilayer on a high-resistance silicon substrate, while the second is based on a NiFe (5 nm)/Pt(5 nm) bilayer deposited on a sapphire substrate. For simplicity, we will in the following refer to these material systems as CoFeB and NiFe, respectively. The SHNO structures are defined by their constriction width ($w$) for single SHNOs and, in array configurations, also by their center-to-center separation ($d$). The nanolithography fabrication process involved depositing blanket film stacks using DC magnetron sputtering in a 3 mTorr argon atmosphere. The stacks were patterned into nano-constrictions using electron beam lithography, followed by ion beam etching. Cu/Pt ground-signal-ground (GSG) contact pads, suitable for a broad radio frequency range, were fabricated using DC sputtering and photolithography. Further details on sample fabrication and their characteristics can be found elsewhere \cite{Zahedinejad2020Two-dimensionalComputing,Kumar2023RobustChains, Kumar2022FabricationNano-oscillators}.

The experimental setup (Fig.~\ref{fig:1 setup}a) employs a two-port vector network analyzer (VNA) to continuously excite the sample with an RF traveling wave ($f_0$) and analyze its optical response. The SHNO is exposed to an external magnetic field $H$ with variable in-plane ($\phi$) and out-of-plane ($\theta$) angles (see Fig.~\ref{fig:1 setup}b). The RF signal from the VNA port 1 is combined with a direct current ($I_{DC}$) through a bias-T and fed into the sample via a GSG picoprobe from GGB Industries. A continuous-wave 532 nm single-mode laser, focused to an approximate 380 nm diameter spot by a microscope objective with a numerical aperture of 0.75, is employed for probing. Both the laser power ($P_{laser}$) and the VNA RF power ($P_{RF}$) can be continuously adjusted within the ranges  0 to 6 mW and $-90$ to 0 dBm, respectively. The sample holder is positioned under the microscope and controlled by an XYZ nanometric stage. The backscattered light from the SHNO passes through an analyzer set at a 45-degree angle relative to the polarization of the probe beam, resulting in an amplitude-modulated beam at the operational frequency of the SHNO. This modulated signal is then converted into an electrical voltage by a fiber-coupled broadband photoreceiver, amplified by 50 dB, and fed to the VNA port 2. For electrical measurements of the free-running SHNO, the signal collected by the coplanar waveguide and the picoprobe is measured using a spectrum analyzer; additional details can be found in~\cite{Kumar2023RobustChains}. All experiments were carried out under room conditions.

The VNA measures the scattering parameter $S_{21} = V_2/V_1$, where $V_1$ is the excitation from port 1 and $V_2$ the output of the photoreceiver after amplification at port 2. The VNA has an intermediate filter (IF) of 10 Hz, sufficient to visualize the autooscillation from a single SHNO structure. To achieve nanometric alignment, we actively stabilized the position of the SHNO using a homemade autofocus system and image recognition software from ThaTEC Innovation.

\begin{figure}[t]
    \centering
    \includegraphics[width=\linewidth]{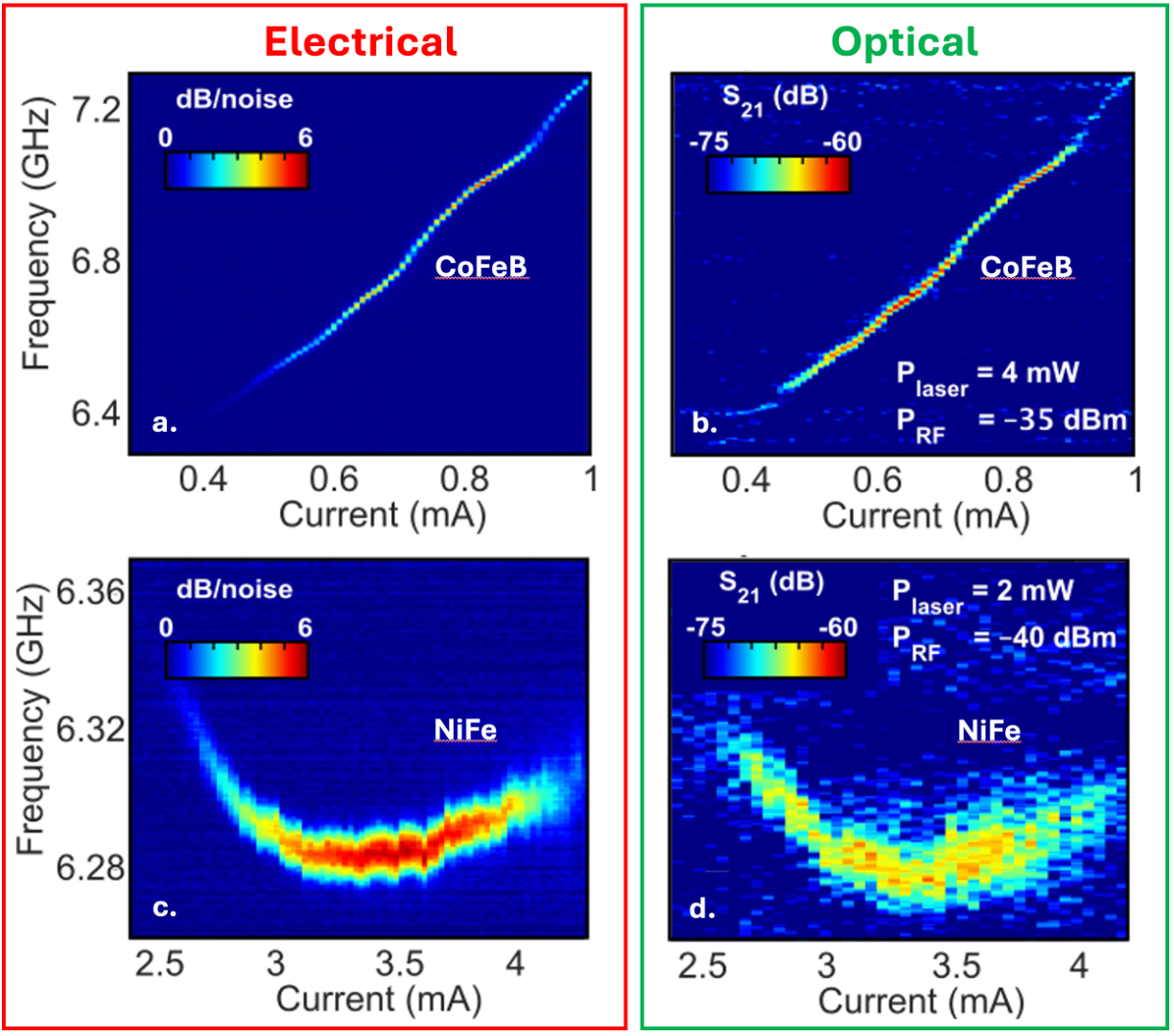}
    \caption{Typical power spectral density (PSD) \emph{vs.}~$I_{DC}$ measurements of CoFeB (a., b.) and NiFe (c., d.) SHNOs. The left column (a., c.) shows traditional electrical measurements and the right column (b., d.) shows the corresponding optical measurements. The applied field conditions were $H=$ 3600 Oe, $\theta=65^\circ$, and $\phi=22.5^\circ$ for CoFeB and $H=$ 6300 Oe, $\theta=82^\circ$, and $\phi=22.5^\circ$ for NiFe.}
    \label{fig:2}
\end{figure}

Figures ~\ref{fig:2}a and \ref{fig:2}b compare the Power Spectral Density (PSD) of a 300 nm wide CoFeB single nanoconstriction obtained through conventional electrical detection (dB over noise) and the optical measurement (in dB) using the VNA $S_{21}$. The SHNO was subject to a $H=$ 3600 Oe magnetic field at angles of $\theta=65^\circ$ and $\phi=22.5^\circ$. For the optical measurement, the laser power was set to $P_{laser}=$ 4 mW, and the RF power to $P_{RF}=$ --35 dBm. Consistent with the typical characteristics of CoFeB SHNOs, a low threshold current of approximately 0.45 mA is extracted together with a positive linear current dependence of the frequency, indicating the excitation of propagating spin waves above the Ferromagnetic Resonance (FMR) frequency, promoted by the perpendicular magnetic anisotropy\cite{Fulara_sciadv2019}.

The corresponding electrical and optical measurements for a 300 nm wide NiFe SHNO, subject to a magnetic field of $H=$ 6300 Oe, $\theta=82^\circ$, and $\phi=22.5^\circ$, and with $P_{laser}=$ 2 mW and $P_{RF}=$ --40 dBm, are presented in Figs.~\ref{fig:2}c,d. Typical for NiFe SHNOs~\cite{Awad2020apl}, the device exhibits a significantly higher threshold current of about 2.5 mA and a nonmonotonic frequency shift that transitions from negative to positive with increasing electrical current. This behavior is characteristic of SHNOs made of materials with in-plane magnetic anisotropy when subject to oblique fields.~\cite{Dvornik_PhysRevApplied2018} In both samples, the optical measurements closely match the features observed with electrical detection. This congruence highlights the reliability and consistency of our optical measurements, demonstrating a strong correspondence with the electrical data.

\begin{figure}[!b]
    \centering
    \includegraphics[width=\linewidth]{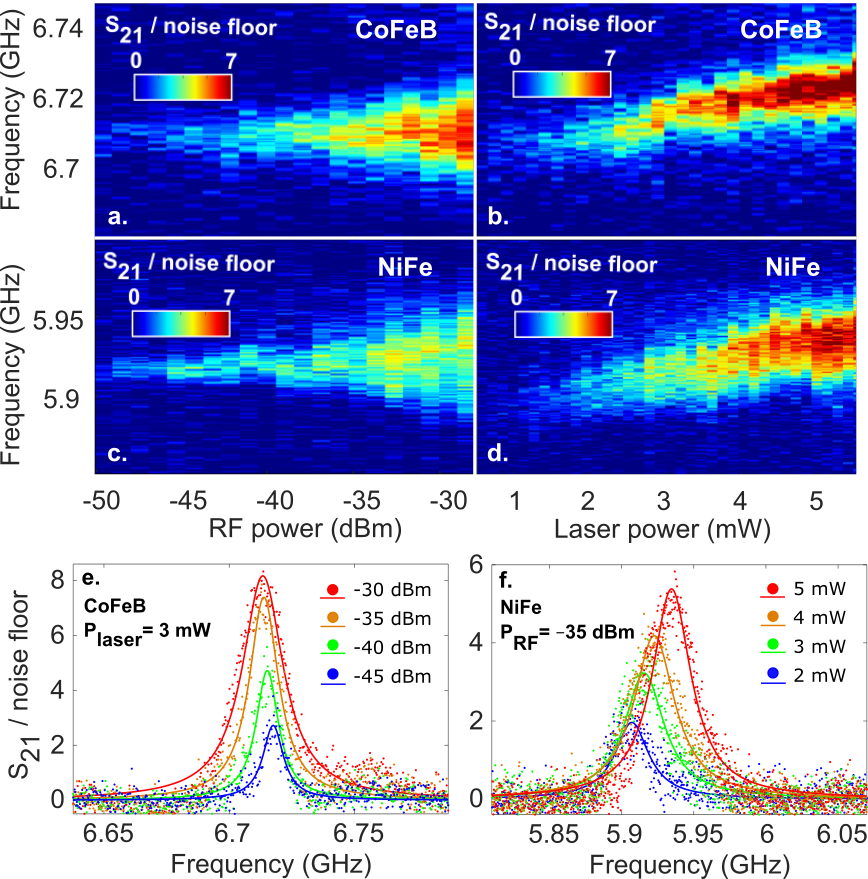}
    \caption{Typical spectra vs $P_{RF}$ and $P_{laser}$ of CoFeB (a., b.) at $I_{DC}=0.65$ mA, and NiFe (c., d.) at $I_{DC}=3.4$ mA. The colorbar scale represents the measured $S_{21}$ normalized over the average noise floor level. The left column shows measurements at fixed $P_{laser}=3$ mW, while the right column measurements at $P_{RF}=-35$ dBm. Two-dimensional spectra with lorentizian fittings (solid lines) are shown for CoFeB (e.) and NiFe (f.).}
    \label{fig:3}
\end{figure}

Electrical PSD measurements are routinely fit with Lorentzian distributions to extract the intrinsic linewidth and total power of the SHNO microwave signal. However, our optical heterodyne method does not strictly measure the intrinsic SHNO signal since it requires the SHNO to be injection-locked to an external reference RF signal. To investigate whether the optical data can still be used for such extraction, we measured the optical spectrum \emph{vs.}~both $P_{RF}$ and $P_{laser}$ (Fig.~\ref{fig:3}) in the hope of extrapolating the measured response to the free-running and unperturbed $P_{RF}=P_{laser}=$ 0.

In the subsequent analysis, the spectrum is presented as the linear magnitude $S_{21}$ normalized to the average value of the noise floor. Starting with the dependence on $P_{RF}$ (Fig.~3a\&c), both the CoFeB and the NiFe SHNOs show the expected simple Arnold tongue behavior for injection locking: The first sign of locking appears at about $P_{RF}=$ --50 dBm, the central frequency of the SHNOs is essentially independent of $P_{RF}$, and the locking bandwidth increases with $P_{RF}$. Fig.~\ref{fig:3}e shows spectra from the CoFeB SHNO, taken at four values of $P_{RF}$, together with Lorentzian fits from which we extract the peak height ($S_0$) and the FWHM width ($\Delta_{VNA}$). The central frequency is essentially independent of the injected power. Fig.~3b\&d show the corresponding PSD of the CoFeB and NiFe SHNOs as a function of $P_{laser}$. We detect the first signal at about 1 mW of laser power. Fig.~\ref{fig:3}f shows example spectra from the NiFe SHNO, taken at four values of $P_{laser}$, together with Lorentzian fits from which we again extract $S_0$ and $\Delta_{VNA}$. In addition to their monotonic increase with $P_{laser}$, the central frequency, too, now shows a clear linear increase with $P_{laser}$. This is a direct effect of the laser heating of the SHNO, which was previously used for optothermal control of the SHNO frequency and will not be discussed further.\cite{Muralidhar2022OptothermalNano-oscillators} 

\begin{figure}
    \centering
    \includegraphics[width=0.82\linewidth]{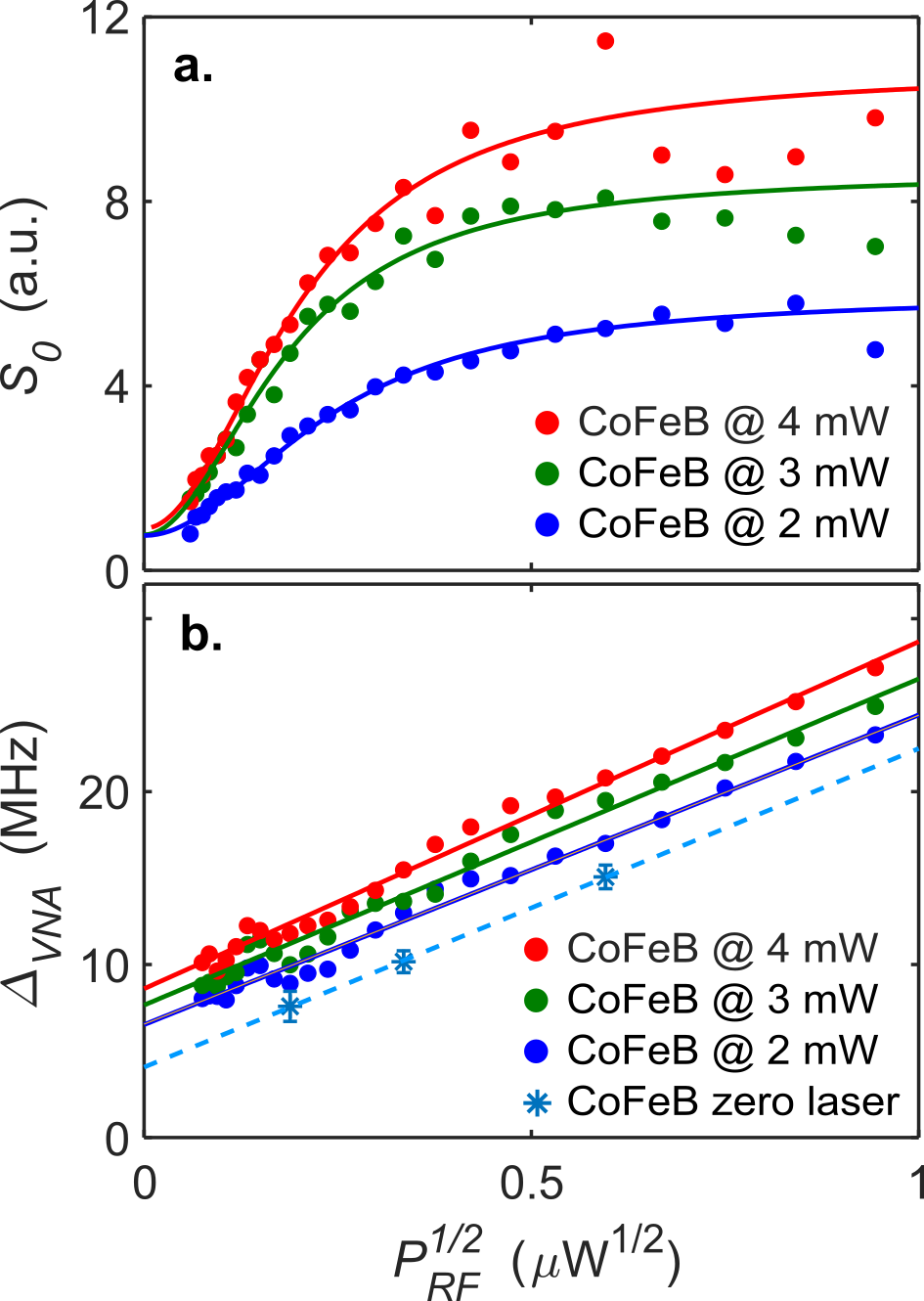}
    \caption{a.~The peak height ($S_0$) and b.~linewidth ($\Delta_{VNA}$) \emph{vs.}~$P_{RF}^{1/2}$, as extracted from Lorentz fits to the normalized $S_{21}$ from the CoFeB SHNO. Solid lines in a.~represent fits to Eq.~1. Solid lines in b. are linear fits \emph{w.r.t.}~$P_{RF}^{1/2}$. The pale blue star symbols denote the zero-laser-power intercepts from fits to the experimental data in Fig.~5b. The dashed blue line is a linear fit. When extrapolated to zero RF power, it yields the intrinsic linewidth ($\Delta f=$ 4.1 MHz) of the free-running SHNO (see text).}
    \label{fig:4}
\end{figure}

We will first analyze and discuss the dependence on the injected RF power. Fig.~\ref{fig:4} plots $S_0$ and $\Delta_{VNA}$ \emph{vs.}~$P_{RF}^{1/2}$ for the CoFeB SHNO at three different values of $P_{laser}$. Plotted against $P_{RF}^{1/2}$, $\Delta_{VNA}$ shows a linear dependence for all SHNOs at all investigated laser powers, confirming the expected square root power dependence of the injection locking bandwidth. $\Delta_{VNA}$ increases slightly with $P_{laser}$, which we will discuss in more detail below. In contrast, $S_0$ shows a non-linear dependence on $P_{RF}^{1/2}$, initially being linear and then asymptotically saturating at a value that increases linearly with $P_{laser}$. This dependence arises from the VNA technique being solely sensitive to the coherent response, \emph{i.e.}~to magnons locked to the injection. Thus, at low injected power levels, both the locking range and the coherent fraction of the magnons are small, but as $P_{RF}$ increases, the coherent fraction grows. However, once the mode fully synchronizes with the RF source, all magnons within the mode oscillate coherently with the injected signal, resulting in the maximum detected amplitude, and any further increases in $P_{RF}$ yield no proportional gains in signal strength. Therefore, the dependence $S_0(P_{RF})$ represents the transformation of the PSD of the noisy oscillator in the vicinity of the injected frequency\cite{chang1997phase,kurokawa1968noise}, which can be rewritten as:
\begin{equation*}
S_0 \propto P_{s}\left(\frac{n_c+a P_{RF}}{a P_{RF}+1}\right),
\end{equation*}
where $n_c$ is a fraction of coherent magnons in a free-running oscillator, and $a$ is a fitting parameter, which depends on the linewidth and power of the free-running oscillator and detection bandwidth. As can be seen from the fits in Fig.~\ref{fig:4}a, Eq.~1 describes the experimental data very well.

\begin{figure}
    \centering
    \includegraphics[width=0.82\linewidth]{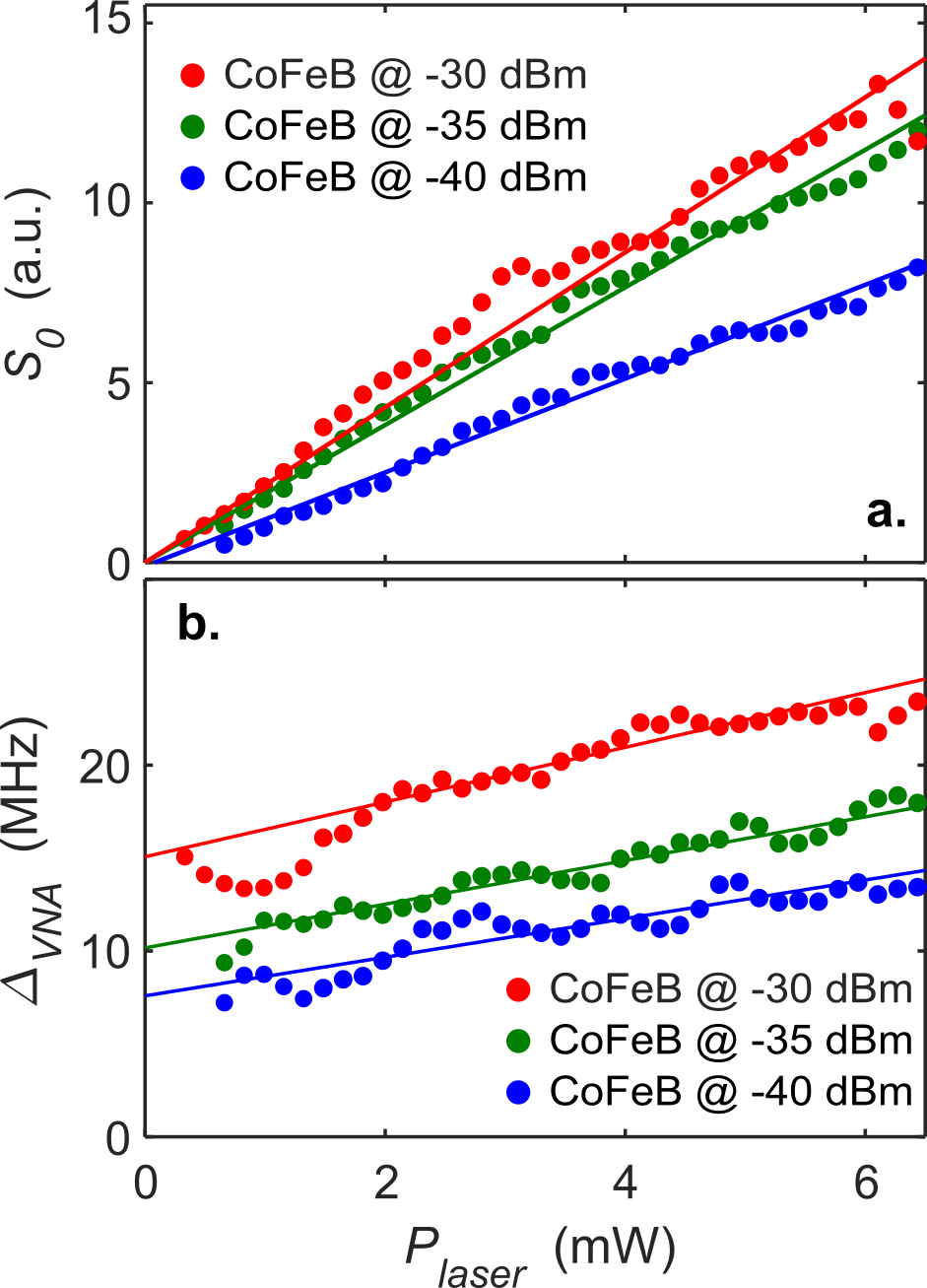}
    \caption{Extracted a.~$S_0$ and b.~$\Delta_{VNA}$ \emph{vs.}~$P_{laser}$ for the CoFeB SHNO. Solid lines represent linear fits to the data. The intercepts with $P_{laser}=$ 0 of the fits in b. are shown as pale blue stars in Fig.~4b and used to extract the intrinsic free-running SHNO linewidth ($\Delta f$).}
    \label{fig:5}
\end{figure}

We now turn to the impact of $P_{laser}$. Fig.~\ref{fig:5} plots $S_0$ and $\Delta_{VNA}$ \emph{vs.}~$P_{laser}$ for the CoFeB SHNO at three different values of $P_{RF}$. $S_0$ is directly proportional to $P_{laser}$, simply reflecting the linear dependence of the photodiode (Fig.~\ref{fig:5}a). There is a slight deviation from the linear behavior towards higher $P_{laser}$, which we ascribe to heating from the laser. The deviation seems to start at lower $P_{laser}$ for stronger injection, which may reflect the combined effect from both optical and electrical heating. 

Since the detection is coherent with the injection signal, the measured $\Delta_{VNA}$ shown in Fig.~\ref{fig:5}b~is approximately the sum of the intrinsic SHNO auto-oscillation linewidth ($\Delta_f$) and the phase-locking bandwidth ($\Delta_{PL}$) to the injected signal, $\Delta_{VNA}\simeq\Delta f+ \Delta_{PL}$. Although the SHNO linewidth is substantially reduced by the injected signal over most of the locking range, it will approach that of the free-running SHNO at the band edges~\cite{chang1997phase}, and therefore add its full intrinsic value to $\Delta_{VNA}$. Both $\Delta_f$ and $\Delta_{PL}$ are inversely dependent on the SHNO oscillation power $p_s$\cite{tiberkevich2014sensitivity}: 
\begin{equation}
\Delta f = \frac{1}{2} \frac{\chi^2}{p_s} S_n(f), \quad \Delta_{PL} = \frac{\chi}{\sqrt{p_s}} P_{rf}^{1/2},
\label{eq:2}
\end{equation}
where $\chi$ is a relative sensitivity of the oscillator to external signals and $S_n(f)$ is the thermal noise power spectrum density, $S_n(f)\propto k_B T$. Hence, when the laser heats up the SHNO, it affects the measured $\Delta_{VNA}$ both through a linear increase of the thermal noise power spectrum, which leads to a broadening of $\Delta f$, and a reduction of the auto-oscillation power $p_s$ (by reducing the magnetization), which increases the locking range $\Delta_{PL}$.

From linear fits to the data in Fig.~\ref{fig:5}b, we can extract the linewidth broadening coefficient due to the laser as $C_{laser}=1.2\pm0.2$ MHz/mW. Extrapolation of these slopes to $P_{laser}=$ 0 then yields $\Delta_{VNA}$ at zero laser power for different values of $P_{RF}$, which is the first step towards extracting the intrinsic SHNO linewidth. In a second step, we add these three extracted values for $\Delta_{VNA}$ to the measured data in Fig.~\ref{fig:4}b and find that they follow the same square root dependence on $P_{RF}$ and with the same slope. A linear fit through these new data points yields broadening coefficients due to the injected RF power of $C_{RF}=17.1\pm2.2$ MHz/$\mu$W$^{1/2}$. Extrapolation to $P_{RF}=$ 0 finally provides us with estimates of the intrinsic SHNO linewidth of $\Delta f=4.1\pm1.6$ MHz, which is consistent with the electrically measured value of 5.9 MHz. Using this measurement and analysis protocol, it is, hence, possible to extract the intrinsic SHNO linewidth through \emph{optical} measurements.

\begin{figure} [!b]
    \centering
    \includegraphics[width=\linewidth]{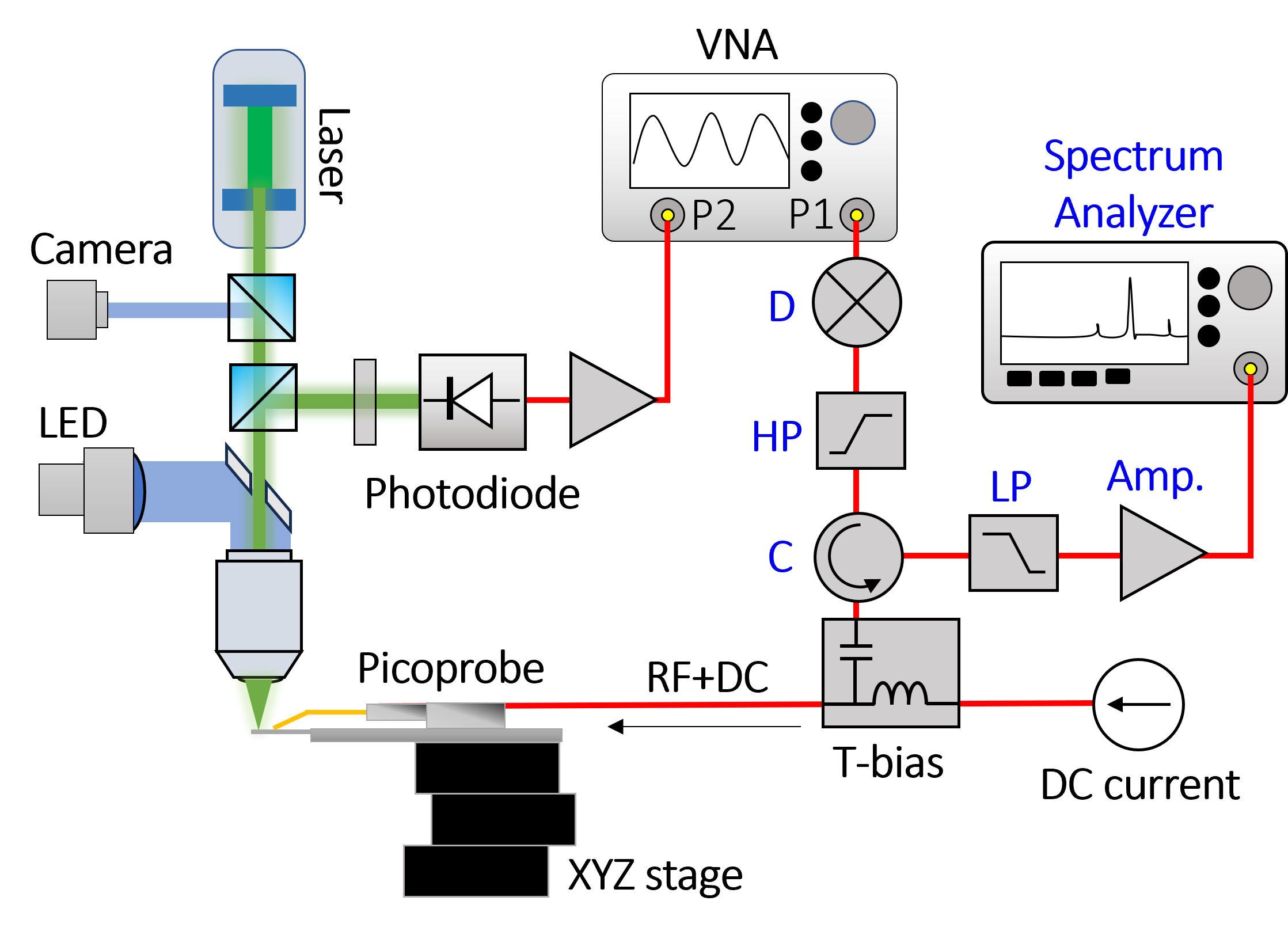}
    \caption{Schematic for parametric injection. D) frequency multiplier, HP) high-pass filter, C) circulator, LP) low-pass filter. Other component labels are referenced as indicated in Fig. \ref{fig:1 setup}.}
    \label{fig:6}
\end{figure}

\begin{figure*}
    \centering   \includegraphics[width=\linewidth]{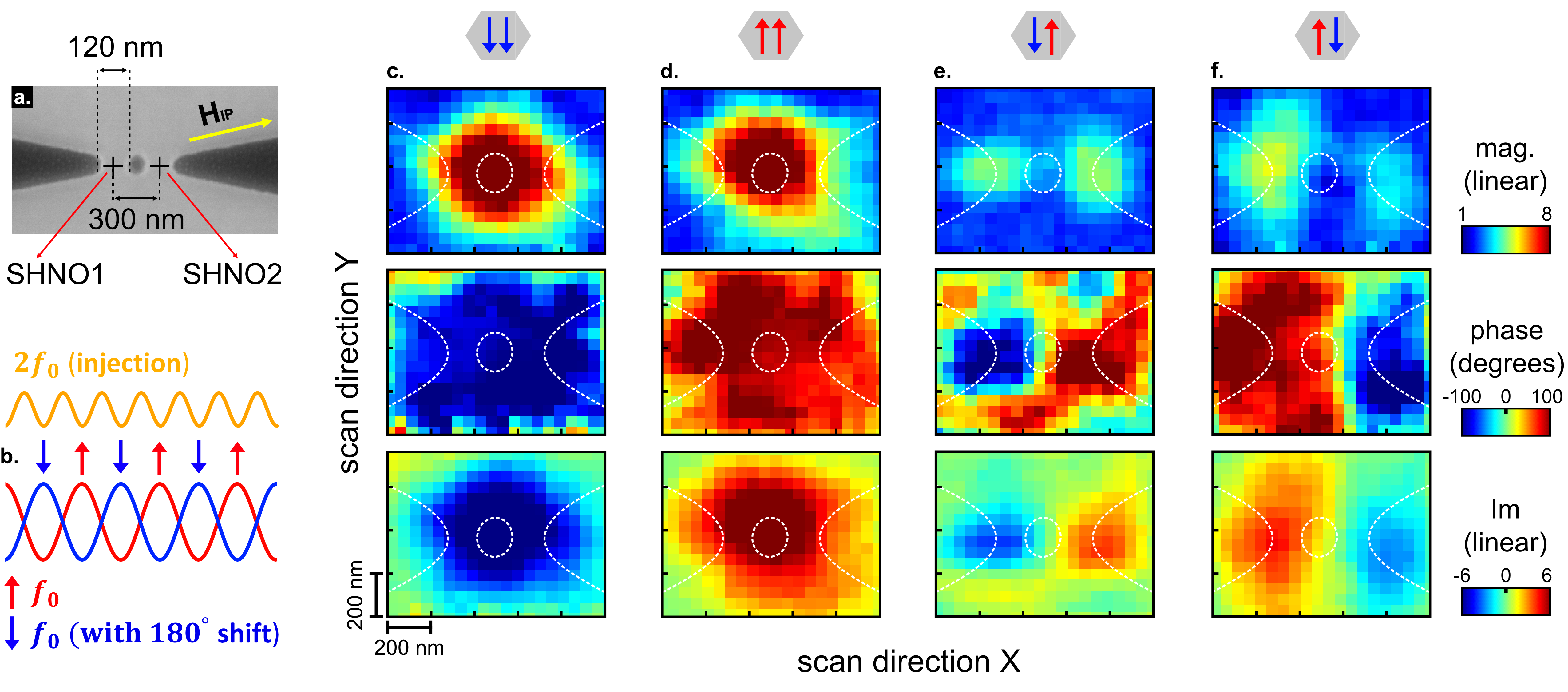}
    \caption{a. SEM image of the SHNO array, $H_{IP}$ represents the in-plane component of the external magnetic field. b. Injection locking scheme at $2f_0$ and the two possible phase states \emph{w.r.t.}~the fundamental frequency $f_0$: red $\langle\uparrow\rangle$ and blue $\langle\downarrow\rangle$ for in-phase and out-of-phase locking. c--f. The four different phase states and spatial maps of the measured VNA signal: The top row shows the $S_{12}$ magnitude, the middle row its phase, and the bottom row its imaginary part. The geometry of the device is outlined by the dotted white lines.}
    \label{fig:7}
\end{figure*}

VNA-based Heterodyne detection accurately captures the auto-oscillation phase, making it particularly advantageous for measuring the relative phase difference between individual SHNOs in SHNO arrays. In the next section, we will demonstrate its usefulness in reading out the individual phase states of two phase-binarized SHNOs of the type used in SHNO Ising machines.\cite{Houshang2022Phase-BinarizedMachines} Phase binarization occurs when oscillators are injection-locked to an external signal at twice their frequency, so-called second harmonic injection locking (SHIL) or parametric pumping. The oscillators can then lock in two degenerate phase states, 0$^\circ$ $\langle\uparrow\rangle$ and 180$^\circ$ $\langle\downarrow\rangle$, \emph{w.r.t}~a reference at the fundamental frequency $f_0$. In our demonstration, we will use a pair of coupled SHNOs, which can have four potential states: $\langle\uparrow\uparrow\rangle$, $\langle\downarrow\downarrow\rangle$, $\langle\uparrow\downarrow\rangle$, and $\langle\downarrow\uparrow\rangle$.

Fig.~\ref{fig:6} illustrates a modification of the original setup where the added electrical components are highlighted in blue text for easy identification while the optical configuration remains the same. The RF frequency from port 1 of the VNA is doubled by a multiplier and then passed through a high-pass filter, which suppresses the original frequency component $f_0$ by 80 dB. The filtered signal is then routed through a circulator into the T-bias/picoprobe circuitry. The electrical signal reflected from the SHNO sample is guided back by the circulator to a spectrum analyzer, passing through a low-pass filter that attenuates the 2$f_0$ component by 80 dB, before being amplified by a 70 dB low noise amplifier.

We conducted scanning imaging on a NiFe/Pt SHNO $2\times 1$ array operating at a frequency of 8.25 GHz, a drive current of 3.75 mA, and a laser power of 2 mW. The sub-harmonic injection power was slowly varied between -10 and 0 dBm until the desired phase state was found and remained stable\cite{Houshang2022Phase-BinarizedMachines}. The array comprises two oscillators with a constriction width of 120 nm and a center-to-center pitch of 300 nm (see Fig.~\ref{fig:6}a). The device was exposed to a magnetic field of 6200 Oe at field angles of $\theta=82^\circ$ and $\phi=22.5^\circ$. The array exhibits robust synchronized states, characterized by well-defined phase-binarized values of 0$^\circ$ $\langle\uparrow\rangle$ and 180$^\circ$ $\langle\downarrow\rangle$ of each oscillator. While for an ideal SHNO array, the phase-binarized states would be pair-wise degenerate in their electrical signal and hence indistinguishable, real SHNOs possess intrinsic biases due to imperfections from the fabrication process, lifting this degeneracy and producing variations in output power for the four states \cite{Houshang2022Phase-BinarizedMachines}. Although determining the 2-dimensional phase state of the device from its electrical output is not feasible a priori, a discrete change in the total auto-oscillation amplitude, monitored by the spectrum analyzer, denotes a state switch in the device. We exploit this characteristic and monitor the electrical microwave signal on the spectrum analyzer to ensure the stability of the system's phase state during the optical measurements. 

Fig. \ref{fig:7} presents spatial maps of the magnitude, phase, and imaginary part of the $S_{21}$ parameter for the four possible states of the SHNO system. The measurement area was $1\times0.8$ $\mu$m with a step resolution of 50 nm, and the VNA integration time per pixel was 102.3 ms, without averaging. The high acquisition speed enables completing the measurement within minutes,  obviating the need for active stabilization of the sample position. The magnitude maps illustrate the mutual phase/antiphase synchronization of the SHNOs, emphasizing the high-power output of the phase-synchronized states. In these states, the individual phases of the two SHNOs are virtually identical, with only the smallest phase difference between them. While the spatial resolution is insufficient to observe the individual SHNOs in the magnitude of the signal, the phase sensitivity of the VNA clearly allows us to conclude that there are two distinct areas with slightly different phases. 

In the antiphase synchronization states, the maps change completely. First, the intensity drops dramatically compared to the in-phase state. Second, the individual oscillators now also become clearly visible in the magnitude maps, as two distinct intensity areas can be observed. Both these dramatic differences are results of coherent destructive wave interference within this region. As the oscillator separation and the laser spot size are about the same, the destructive interference affects the entire map. However, the cancellation is only complete half-way between the oscillators, which leaves partial intensities at the oscillator locations. Further insights are gained by examining the phase phase information in the $S_{21}$ signal, clearly revealing the relative anti-phase state of the two oscillators. In addition, the phase map shows a remarkable spatial resolution of about 100--150 nm, which is much below the diffraction limit. The phase maps extend well outside the auto-oscillating regions, where it is strictly speaking not well defined since the magnitude of $S_{21}$ falls into the noise. It may therefore be more useful to look at the imaginary part of $S_{21}$, which simultaneously show both the intensity and the phase state of the auto-oscillating regions.

In this specific measurement, we calculated a contrast of approximately 31 dB between the average noise floor and the maximum of the imaginary component of $S_{21}$.

\section*{Conclusions}

We demonstrate a robust optical heterodyne detection microscopy technique for characterizing the magnetic auto-oscillations of nano-constriction based Spin Hall nano-oscillators. The detailed characterization of both single and interacting SHNOs made from two distinctly different material systems highlights the technique's robustness and its potential for wider applications. Our heterodyne detection scheme enables the measurement of key parameters of the SHNOs, including their frequency, amplitude, locking bandwidth to externally injected rf signal, and response to optothermal tuning. Notably, the technique allows for the effective de-embedding of perturbing effects from the injected power and the laser heating, providing optical estimates of the SHNO linewidth even in its free-running state. The technique offers significant advantages over alternative methods optical methods. For instance, our approach surpasses the frequency resolution limits of Brillouin Light Scattering measurements, which is typically constrained by the Tandem Fabry-Pérot Interferometer's resolution of around 100 MHz. In contrast, our method benefits from the Vector Network Analyzer's (VNA) orders of magnitude better resolution, typically below the kHz range. Moreover, unlike interferometry, cavity-based spectrometers that necessitate continuous scanning and stabilization of the reference wavelength, our optical heterodyne technique facilitates continuous wave acquisition, significantly speeding up the acquisition process by orders of magnitude. Finally, a key feature of this technique is the very sensitive phase-resolved characterization of interacting SHNOs, crucial for using phase-binarized SHNOs in Ising Machines, where the technique easily resolves the phase states of SHNOs separated by 300 nm, with a spatial resolution of 150 nm.

\section*{Author Contributions}
\textbf{A. Aleman:} Instrumentation design \& construction (lead), experimental design (equal), development \& design of methodology (lead), software development (lead), data acquisition (lead), data curation (equal), formal analysis (lead), original draft preparation (lead), writing-reviewing \& editing (equal). \textbf{A. A. Awad:} Conceptualization and original idea (lead), experimental design (equal), data curation (equal), research planning (equal), writing-reviewing \& editing (equal). \textbf{S. Muralidhar} Software development (supporting), investigation (supporting). \textbf{R. Khymyn} Theoretical analysis (lead), critical review (supporting). \textbf{A. Kumar} Sample preparation (equal), data acquisition (supporting). \textbf{A. Houshang:} Sample preparation (equal). \textbf{D. Hanstorp:} project supervision (supporting), funding acquisition (supporting), critical review (supporting). \textbf{J. \AA kerman} Project supervision \& management (lead), research planning (equal), data curation (equal), writing-review \& editing (lead), funding acquisition (lead), critical review (lead).      
\section*{Conflicts of interest}
There are no conflicts to declare
\section*{Acknowledgements}
This work was partially supported by the Horizon 2020 research and innovation programmes No.~835068 "TOPSPIN" and No.~899559 "SpinAge", the Swedish Research Council (Grant No. 2016-05980), and the Knut and Alice Wallenberg Foundation.



\balance


\bibliography{rsc} 
\bibliographystyle{rsc} 

\end{document}